\documentclass{iau} 
\usepackage{graphicx}
\usepackage{natbib}
\usepackage{url}
\bibpunct{(}{)}{;}{a}{}{,}

\def\gtsim {>\kern-1.2em\lower1.1ex\hbox{$\sim$}~}   % Greater than sim
\def\ltsim {<\kern-1.2em\lower1.1ex\hbox{$\sim$}~}   % Less than sim
\def \apj {\textit{ApJ}}
\def \apjl {\textit{ApJ}, Letter}
\def \apjs {\textit{ApJS}}

\def \aap {\textit{A\&A}}
\def \mnras {\textit{MNRAS}}

\def \araa {\textit{ARAA}}
\def \pasa {\textit{PASA}}
\def \nat {\textit{Nature}}

\title[IAU 366.~~The role of mass loss] %% give here short title %%
{The role of mass loss in chemodynamical evolution of galaxies}

\author[Chiaki Kobayashi]   %% give here short author list %%
{Chiaki Kobayashi$^1$}

\affiliation{$^1$
Centre for Astrophysics Research,
Department of Physics, Astronomy and Mathematics
University of Hertfordshire,
College Lane, Hatfield  AL10 9AB, UK
email: {\tt c.kobayashi@herts.ac.uk}}

\pubyear{2022}
\volume{366}  %% insert here IAU Symposium No.
\jname{The origin of outflows in evolved stars}
\editors{Leen Decin, Albert Zijlstra, Clio Gielen}
\begin{document}

\maketitle

\begin{abstract}
Thanks to the long-term collaborations between nuclear and astrophysics, we have good understanding on the origin of elements in the universe, except for the elements around Ti and some neutron-capture elements.
From the comparison between observations of nearby stars and Galactic chemical evolution models, a rapid neutron-capture process associated with core-collapse supernovae is required.
The production of C, N, F and some minor isotopes depends on the rotation of massive stars, and the observations of distant galaxies with ALMA indicate rapid cosmic enrichment.
It might be hard to find very metal-poor or Population III (and dust-free) galaxies at very high redshifts even with JWST.
\end{abstract}

\firstsection % if your document starts with a section,
\vspace*{-2mm}
\section{Introduction}

\begin{figure}[t]
\begin{center}
\includegraphics[width=7cm]{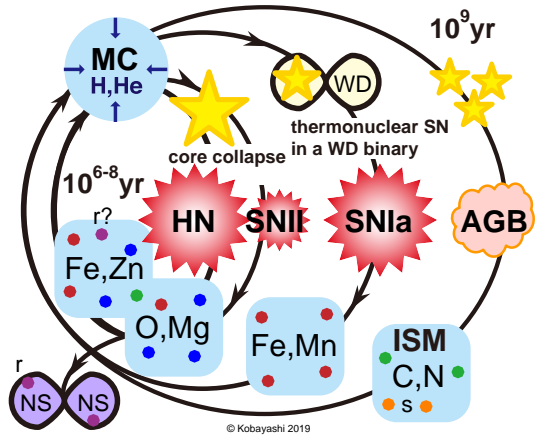}
\caption{\label{fig:intro}
Schematic view of chemical enrichment in galaxies.
}
\end{center}
\end{figure}

Explaining the origin of the elements is one of the scientific triumphs linking nuclear physics with astrophysics. As Fred Hoyle predicted, carbon and heavier elements (`metals' in astrophysics) were not produced during the Big Bang but instead created inside stars. So-called $\alpha$ elements (O, Ne, Mg, Si, S, Ar, and Ca) are mainly produced by core-collapse supernovae, while iron-peak elements (Cr, Mn, Fe, and Ni) are more produced by thermonuclear explosions, observed as Type Ia supernovae (SNe Ia; \citealt{kob20sr}, hereafter K20). The production depends on the mass of white dwarf (WD) progenitors, and a large fraction of SNe Ia should come from near-Chandrasekhar (Ch) mass explosions (see \citealt{kob20ia} for constraining the relative contribution between near-Ch and sub-Ch mass SNe Ia).
Among core-collapse supernovae, hypernovae ($\gtsim 10^{52}$ erg) produce a significant amount of Fe as well as Co and Zn, and a significant fraction of massive stars ($\gtsim 20M_\odot$) should explode as hypernovae in order to explain the Galactic chemical evolution (GCE; \citealt{kob06}, hereafter K06).

Heavier elements are produced by neutron-capture processes. The slow neutron-capture process (s-process) occurs in asymptotic giant branch (AGB) stars \citep{kar16}, while the astronomical sites of rapid neutron-capture process (r-process) have been debated. The possible sites are neutron-star (NS) mergers \citep[NSM,][]{wan14,jus15}, magneto-rotational supernovae \citep[MRSNe,][]{nis15,rei21}, magneto-rotational hypernovae/collapsars \citep[MRHNe,][]{yon21a}, and common envelope jets supernovae \citep{gri22}. Light neutron-capture elements (e.g., Sr) are also produced by electron-capture supernovae (ECSNe, \citealt{wan13ecsn}), $\nu$-driven winds \citep{arc07,wan13nu}, and rotating-massive stars \citep{fri16,lim18}.

The cycles of chemical enrichment are schematically shown in Figure \ref{fig:intro}, where each cycle produces different elements and isotopes with different timescales.
In a galaxy, not only the total amount of metals, i.e. metallicity $Z$, but also elemental abundance ratios evolve as a function of time. 
Therefore, we can use all of this information as fossils to study the formation and evolutionary histories of the galaxy. This approach is called Galactic archaeology, and several on-going and future surveys with multi-object spectrographs (e.g., APOGEE, HERMES-GALAH, Gaia-ESO, WEAVE, 4MOST, MOONS, Subaru Prime Focus Spectrograph (PFS), and Maunakea Spectroscopic Explorer (MSE)) are producing a vast amount of observational data of elemental abundances.
Moreover, integral field unit (IFU) spectrographs (e.g., SAURON, SINFONI, CALIFA, SAMI, MaNGA, KMOS, MUSE, and HECTOR) allow us to measure metallicity and elemental abundance ratios within galaxies. It is now possible to apply the same approach not only to our own Milky Way but also to other types of galaxies or distant galaxies. Let us call this extra-galactic archaeology.

One of the most important uncertainties in this quest is the input stellar physics, namely, the nucleosynthesis yields, which could directly affect conclusions on the formation and evolutionary histories of galaxies.
In this review, I will focus on the impact of stellar mass loss due to stellar rotation in Galactic and extra-galactic archaeology.

\vspace*{-2mm}
\section{Galactic chemical evolution}
\label{sec:gce}

Galactic chemical evolution (GCE) has been calculated analytically and numerically using the following equation:
\begin{equation}\label{eq:gce}
\frac{d(Z_if_{\rm g})}{dt}=E_{\rm SW}+E_{\rm SNcc}+E_{\rm SNIa}+E_{\rm NSM}-Z_i\phi+Z_{i,{\rm inflow}}R_{\rm inflow}-Z_iR_{\rm outflow}
\end{equation}
where the mass fraction of each element $i$ in gas-phase ($f_{\rm g}$ denotes the gas fraction) increases via element ejections from stellar winds ($E_{\rm SW}$), core-collapse supernovae ($E_{\rm SNcc}$), Type Ia supernovae ($E_{\rm SNIa}$), and neutron star mergers ($E_{\rm NSM}$). It also decreases by star formation (with a rate $\phi$), and modified by inflow (with a rate $R_{\rm inflow}$) and outflow (with a rate $R_{\rm outflow}$) of gas in/from the system considered.
It is assumed that the elemental abundance of gas is instantaneously well mixed in the system (called an one-zone model), but the instantaneous recycling approximation is not adopted nowadays. %\citep{tin80,mat21}.
The initial conditions are $f_{{\rm g},0}=1$ (a closed system) or $f_{{\rm g},0}=0$ (an open system) with the chemical composition ($Z_{i,0}$) from the Big Bang nucleosynthesis.
The first two terms depend only on nucleosynthesis yields, while the third and fourth terms also depend on modelling of the progenitor systems, which is uncertain. The last three terms are galactic terms, and should be determined from galactic dynamics, but are assumed with analytic formula in GCE models.

As in my previous works, star formation rate is assumed to be proportional to gas fraction as $\phi\!=\!\frac{1}{\tau_{\rm s}}f_{\rm g}$. The inflow rate is assumed to be $R_{\rm inflow}\!\!=\!\frac{t}{\tau_{\rm i}^2}\exp\frac{-t}{\tau_{\rm i}}$ (for the solar neighbourhood), or $R_{\rm inflow}\!\!=\!\frac{1}{\tau_{\rm i}}\exp\frac{-t}{\tau_{\rm i}}$ (for the rest). The outflow rate is assumed to be proportional to the star formation rate as $R_{\rm outflow}\!\!=\!\frac{1}{\tau_{\rm o}}f_{\rm g}\propto\phi$. In addition, star formation is quenched ($\phi=0$) in case with a galactic wind at an epoch $t_{\rm w}$.
The timescales are determined to match the observed metallicity distribution function (MDF) in each system (Fig.\,2 of K20). The parameter sets that have very similar MDFs give almost identical tracks of elemental abundance ratios (Fig.\,A1 of \citealt{kob20ia}).

Our nucleosynthesis yields of core-collapse supernovae were originally calculated in K06, 3 models of which are replaced in \citet{kob11agb} (which are also used in \citealt{nom13}), and a new set with failed supernovae is used in K20. This is based on the lack of observed progenitors at supernova locations in the HST data \citep{sma09}, and on the lack of successful explosion simulations for massive stars \citep{jan12,bur21}.
The yields for AGB stars were originally calculated in \citet{kar10} and \citet{kar16}, but a new set with the s-process is used in K20. The narrow mass range of super-AGB stars is also filled with the yields from \citet{doh14a}; stars at the massive end are likely to become ECSNe as Crab Nebula exploded in 1054. At the low-mass end, off-centre ignition of C flame moves inward but does not reach the centre, which remains a hybrid C+O+Ne WD. This might become a sub-class of SNe Ia \citep{kob15}.
For SNe Ia, the nucleosynthesis yields of near-Ch and sub-Ch mass models are newly calculated in \citet{kob20ia}, which used the same code as in \citet{leu18} and \citet{leu20} but with more realistic, solar-scaled initial composition. The initial composition gives significantly different (Ni, Mn)/Fe ratios.
r-process yields are taken from literature (see \S 2.1 of K20 for more details).
The Kroupa initial mass function (IMF) between $0.01M_\odot$ and $50M_\odot$ (or $120M_\odot$ in \S \ref{sec:f}, \ref{sec:iso}, and \S \ref{sec:hydro}) is used throughout this paper.
For comparison to observational data, the solar abundances are taken from \citet{asp09}, except for $A_\odot({\rm O}) = 8.76,  A_\odot({\rm Th}) = 0.22$, and $A_\odot({\rm U}) = -0.02$ (\S 2.2 of K20 for the details).

\begin{figure*}[t]
\begin{center}
  \includegraphics[width=0.95\textwidth]{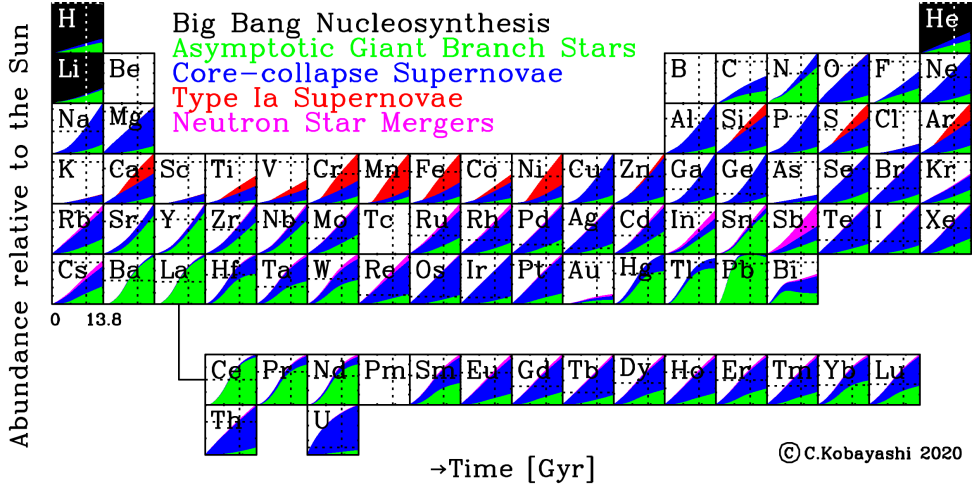}
\caption{The time evolution (in Gyr) of the origin of elements in the periodic table: Big Bang nucleosynthesis (black), AGB stars (green), core-collapse supernovae including SNe II, HNe, ECSNe, and MRSNe (blue), SNe Ia (red), and NSMs (magenta). The amounts returned via stellar mass loss are also included for AGB stars and core-collapse supernovae depending on the progenitor mass. The dotted lines indicate the observed solar values.}
\label{fig:origin}
\end{center}
\end{figure*}

Using the K20 GCE model for the solar neighbourhood, we summarize the origin of elements in the form of a periodic table. In each box of Figure \ref{fig:origin}, the contribution from each chemical enrichment source is plotted as a function of time: Big Bang nucleosynthesis (black), AGB stars (green), core-collapse supernovae including SNe II, HNe, ECSNe, and MRSNe (blue), SNe Ia (red), and NSMs (magenta).
It is important to note that the amounts returned via stellar mass loss are also included for AGB stars and core-collapse supernovae depending on the progenitor star mass.
The x-axis of each box shows time from $t=0$ (Big Bang) to $13.8$ Gyrs, while the y-axis shows the linear abundance relative to the Sun, $X/X_\odot$.
The dotted lines indicate the observed solar values, i.e., $X/X_\odot=1$ and $4.6$ Gyr for the age of the Sun.
Since the Sun is slightly more metal-rich than the other stars in the solar neighborhood (see Fig.\,2 of K20), the fiducial model goes through [O/Fe]$=$[Fe/H]$=0$ slightly later compared with the Sun's age.
Thus, a slightly faster star formation timescale ($\tau_{\rm s}=4$ Gyr instead of 4.7 Gyr) is adopted in this model.
The evolutionary tracks of [X/Fe] are almost identical.
The adopted star formation history is similar to the observed cosmic star formation rate history, and thus this figure can also be interpreted as the origin of elements in the universe.
Note that Tc and Pm are radioactive.
The origin of stable elements can be summarized as follows:
\begin{itemize}
\item H and most of He are produced in Big Bang nucleosynthesis. As noted, the green and blue areas also include the amounts returned to the interstellar medium (ISM) via stellar mass loss in addition to He newly synthesized in stars.
Be and B are supposed to be produced by cosmic rays, %\citep{pra93}, 
which are not included in the K20 model.
\item The Li model is very uncertain because the initial abundance and nucleosynthesis yields are uncertain. Li is supposed to be produced also by cosmic rays and novae, which are not included in the K20 model. The observed Li abundances show an increasing trend from very low metallicities to the solar metallicity, which could be explained by cosmic rays. Then the observation shows a decreasing trend from the solar metallicities to the super-solar metallicities, which might be caused by the reduction of the nova rate \citep{gri19}; this is also shown in theoretical calculation with binary population synthesis \citep{kem22}, where the nova rate becomes higher due to smaller stellar radii and higher remnant masses at low metallicities.
\item 49\% of C, 51\% of F, and 74\% of N are produced by AGB stars (at $t=9.2$ Gyr). Note that extra production from Wolf-Rayet (WR) stars is not included and may be important for F (see \S \ref{sec:f}).
For the elements from Ne to Ge, the newly synthesized amounts are very small for AGB stars, and the small green areas are mostly for mass loss.
\item $\alpha$ elements (O, Ne, Mg, Si, S, Ar, and Ca) are mainly produced by core-collapse supernovae, but 22\% of Si, 29\% of S, 34\% of Ar, and 39\% of Ca come from SNe Ia. 
These fractions would become higher with sub-Ch-mass SNe Ia \citep{kob20ia} instead of 100\% Ch-mass SNe Ia adopted in the K20 model.
\item A large fraction of Cr, Mn, Fe, and Ni are produced by SNe Ia. In classical works, most of Fe was thought to be produced by SNe Ia, but the fraction is only 60\% in our model, and the rest is mainly produced by HNe. The inclusion of HNe is very important as it changes the cooling and star formation histories of the universe significantly \citep{kob07}.
Co, Cu, Zn, Ga, and Ge are largely produced by HNe.
In the K20 model, 50\% of stars at $\ge 20M_\odot$ are assumed to explode as hypernovae, and the rest of stars at $> 30M_\odot$ become failed supernovae.
\item Among neutron-capture elements, as predicted from nucleosynthesis yields, AGB stars are the main enrichment source for the s-process elements at the second (Ba) and third (Pb) peaks. 
\item 32\% of Sr, 22\% of Y, and 44\% of Zr can be produced from ECSNe, which are included in the blue areas, even with our conservative mass ranges; we take the metallicity-dependent mass ranges from the theoretical calculation of super-AGB stars \citep{doh15}.
Combined with the contributions from AGB stars, it is possible to perfectly reproduce the observed trends, and no extra light element primary process (LEPP) is needed. The inclusion of $\nu$-driven winds in GCE simulation results in a strong overproduction of the abundances of the elements from Sr to Sn with respect to the observations.
\item For the heavier neutron-capture elements, contributions from both NS-NS/NS-black hole (BH) mergers and MRSNe are necessary, and the latter is included in the blue areas.
\end{itemize}

In this model, the O and Fe abundances go though the cross of the dotted lines, meaning [O/Fe] $=$ [Fe/H] $=0$ at 4.6 Gyr ago.
This is also the case for some important elements including N, Ca, Cr, Mn, Ni, Zn, Eu, and Th.
Mg is slightly under-produced in the model, although the model gives a $0.2-0.3$ dex higher [Mg/Fe] value than observed at low metallicities.
This [O/Mg] problem is probably due to uncertain nuclear reaction rates (namely, of $^{12}$C($\alpha$,$\gamma$)$^{16}$O) and/or the mass loss based by stellar rotation or binary interaction (see also Fig.\,9 of K06).

The under-production of the elements around Ti is a long-standing problem, which was shown to be enhanced by multi-dimensional effects (\citealt{sne16}; see also K15 model in K20).
The s-process elements are slightly overproduced even with the updated s-process yields.
Notably, Ag is over-produced by a factor of $6$, while Au is under-produced by a factor of $5$. U is also over-produced. These problems may require revising nuclear physics modelling (see \S \ref{sec:au}).

\begin{figure}[t]
\begin{center}
  \includegraphics[width=0.6\textwidth]{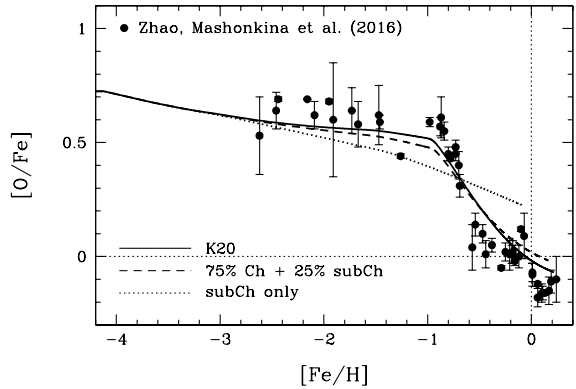}
\vspace*{-2mm}
\caption{The [O/Fe]--[Fe/H] relations in the solar neighborhood for the models with Ch-mass SNe Ia only (solid line), 75\% Ch plus 25\% sub-Ch-mass SNe Ia (dashed line), and sub-Ch-mass SNe Ia only (dotted line). The observational data (filled circles) are high-resolution non-local thermodynamic equilibrium (NLTE) abundances.
}
\label{fig:ofe}
\end{center}
\end{figure}

The contributions from SNe Ia depend on the mass of the progenitor WDs, and sub-Ch mass explosions produce less Mn and Ni, and more Si, S, and Ar than near-Ch mass explosions.
Figure \ref{fig:ofe} shows the [O/Fe]--[Fe/H] relations with varying the fraction of sub-Ch-mass SNe Ia. Including up to 25\% sub-Ch mass contribution to the GCE (dashed line) gives a similar relation as the K20 model (solid line), while the model with 100\% sub-Ch-mass SNe Ia (dotted line) gives too low an [O/Fe] ratio compared with the observational data. For Ch-mass SNe Ia, the progenitor model is based on the single-degenerate scenario with the metallicity effect due to optically thick winds \citep{kob98}. For sub-Ch-mass SNe Ia, the observed delay-time distribution is used since the progenitors are the combination of mergers of two WDs in double degenerate systems and low accretion in single degenerate systems; \citet{kob15}'s formula are for those in single degenerate systems only.

The [$\alpha$/Fe]--[Fe/H] relation is probably the most important diagram in GCE.
In the beginning of the universe, the first stars form and die, but the properties such as mass and rotation are uncertain and have been studied using the second generation, extremely metal-poor (EMP) stars.
Secondly, core-collapse supernovae occur, and their yields are imprinted in the Population II stars in the Galactic halo. The [$\alpha$/Fe] ratio is high and stays roughly constant with a small scatter. This plateau value does not depend on the star formation history but does on the IMF.
Finally, SNe Ia occur, which produce more Fe than O, and thus the [$\alpha$/Fe] ratio decreases toward higher metallicities; this decreasing trend is seen for the Population I stars in the Galactic disk.
Because of this [$\alpha$/Fe]--[Fe/H] relation, high-$\alpha$ and low-$\alpha$ are often used as a proxy of old and young age of stars in galaxies, respectively. Note that, however, that this relation is not linear but is a plateau and decreasing trend (called a `knee'). The location of the knee depends on the star formation timescale and is at a high metallicity in the Galactic bulge, followed by the Galactic thick disk, thin disk, and satellite galaxies show at lower metallicities.

\subsection{Mystery of gold}
\label{sec:au}

\begin{figure*}
\center
  \includegraphics[width=0.95\textwidth]{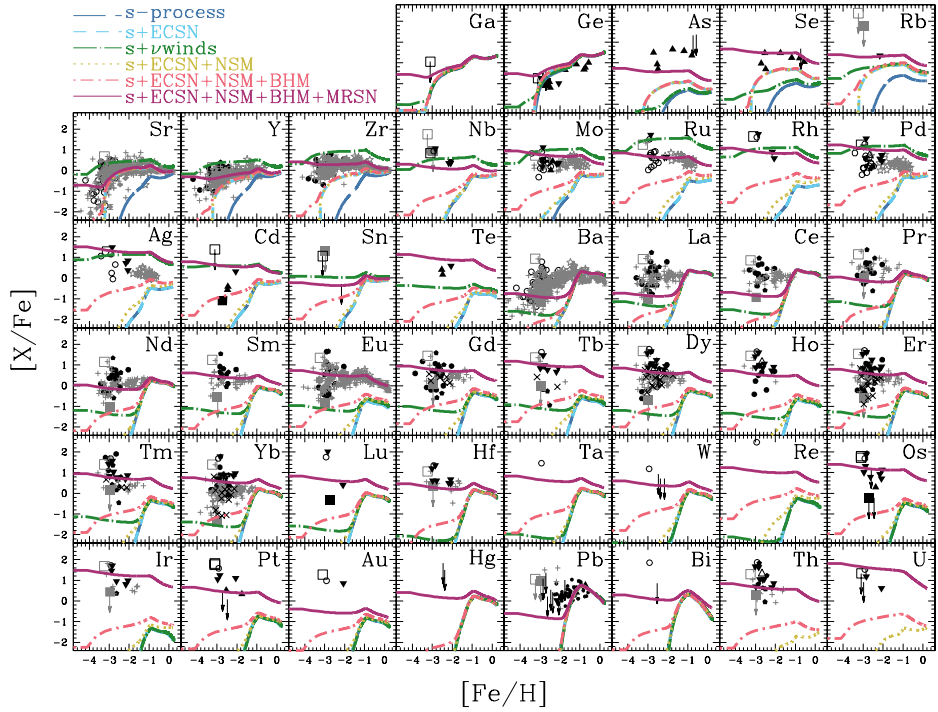}
\caption{The [X/Fe]--[Fe/H] relations for neutron capture elements, comparing to the models in the solar neighborhood, with s-process from AGB stars only (blue long-dashed lines), with s-process and ECSNe (light-blue short-dashed lines), with s-process, ECSNe, and $\nu$-driven winds (green dotted-long-dashed lines), with s-process, ECSNe, and NS-NS mergers (olive dotted lines), with s-process, ECSNe, and NS-NS/NS-BH mergers (orange dotted-short-dashed lines), with s-process, ECSNe, NS-NS/NS-BH mergers, and MRSNe (red solid lines). 
Observational data are updated from K20.}
\label{fig:xfe}       % Give a unique label
\end{figure*}

Figure \ref{fig:xfe} shows the evolutions of neutron-capture elements as [X/Fe]--[Fe/H] relations. 
As known, AGB stars can produce the first (Sr, Y, Zr), second (Ba), and third (Pb) peak s-process elements, but no heaver elements (navy long-dashed lines). It is surprising that ECSNe from a narrow mass range ($\Delta M \sim 0.15-0.2M_\odot$) can produce enough of the first peak elements; with the combination of AGB stars, it is possible to reproduce the observational data very well (cyan short-dashed lines). This means that no other light element primary process (LEPP), such as rotating massive stars, is required. However, the elements from Mo to Ag seem to be overproduced, which could be tested with the UV spectrograph proposed for VLT, CUBES.  Additional production from $\nu$-driven winds leads to further over-production of these elements in the model (green dot-long-dashed lines), but this should be studied with more self-consistent calculations of supernova explosions.

Neutron star mergers can produce lanthanides and actinides, but not enough (olive dotted lines); the rate is too low and the timescale is too long, according to binary population synthesis. This is not improved enough even if neutron-star black-hole mergers are included (orange dot-short-dashed lines). An r-process associated with core-collapse supernovae, such as MRSNe, is required. The same conclusion is obtained with other GCE models and more sophisticated chemodynamical simulations \citep[e.g.,][]{hay19}, as well as from the observational constraints of radioactive nuclei in the solar system \citep{wal21}.

In the GCE model with MRSNe (magenta solid lines), it is possible to reproduce a plateau at low metallicities for Eu, Pt, and Th, relative to Fe. However, even with including both MRSNe and neutron star mergers, the predicted Au abundance is more than ten times lower than observed. This underproduction is seen not only for the solar abundance but also for low metallicity stars although the observational data are very limited. UV spectroscopy is needed for investigating this problem further.

In the GCE models, we adopt the best available nucleosynthesis yields in order to explain the universal r-process pattern: $25M_\odot$ ``b11tw1.00'' 3D yields from \citet{nis15} and $1.3M_\odot$+$1.3M_\odot$ 3D/GR yields from \citet{wan14}.
These yields are sensitive to the electron fraction, which depends on hydrodynamics and $\nu$-processes during explosions. Post-process nucleosynthesis calculations of successful explosion simulations of massive stars are required.
However, it would not be easy to increase Au yields only since Pt is already in good agreement with the current model. There are uncertainties in some nuclear reaction rates and in the modelling of fission, %\citep{shi16,vas20}, 
which might be able to increase Au yields only, without increasing Pt or Ag.
The predicted Th and U abundances are after the long-term decay, to be compared with observations of metal-poor stars, and the current model does not reproduce the Th/U ratio either.

It seems necessary to have the r-process associated with core-collapse supernovae, such as MRSNe.
Is there any direct evidence for such events?
There are a few magneto-hydrodynamical simulations and post-process nucleosynthesis that successfully showed enough neutron-rich ejecta to produce the 3rd peak r-process elements \citep{win12,mos18} as in the Sun. However, there is no observational evidence that support magneto-rotational supernovae. The predicted iron mass is rather small \citep{nis15,rei21}, and the astronomical object could be faint.

\begin{figure}[t]
\begin{center}
  \includegraphics[width=0.65\textwidth]{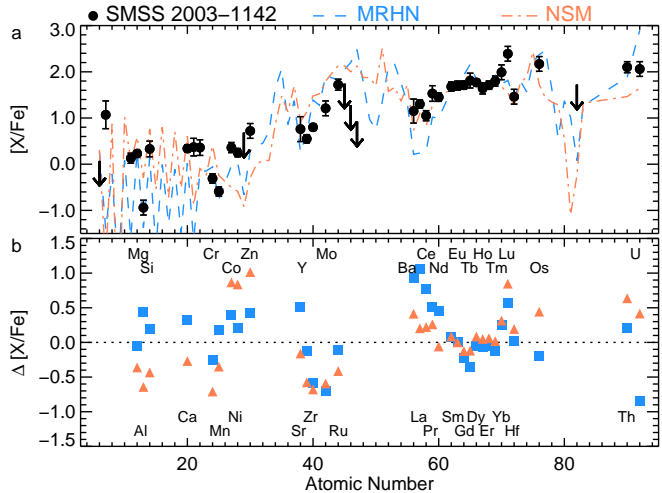}
\caption{The elemental abundance of a extremely metal-poor star SMSS J200322.54-114203.3, which has [Fe/H]$= -3.5$, comparing with mono-enrichment from a magneto-rotational hypernovae (blue dashed line and squares) and multi-enrichment including a neutron star merger (orange dot-dashed line and triangles).
The lower panel shows the differences, i.e., the observed values minus model values.
}
\label{fig:yong}
\end{center}
\end{figure}

Extremely metal-poor (EMP) stars have been an extremely useful tool in the Galactic archeology. At the beginning of galaxy formation, stars form from a gas cloud that was enriched only by one or very small number of supernovae, and hence the elemental abundances of EMP stars have offered observational evidences of supernovae in the past.
As a result, it is found that quite a large fraction of massive stars became faint supernovae that give a high C/Fe ratio leaving a relatively large black hole ($\sim 5M_\odot$, \citealt{nom13}). It might even possible to form a black hole from $10-20M_\odot$ stars \citep{kob14}.
It is challenging to find EMP stars due to the expected small number, and it is also challenging to measure elemental abundances in detail. Thus only a limited number of EMP stars have been analysed in previous works. %\citep{cay04,hon04}.
The Australian team has been using a strategic approach to increase EMP data; $\sim 26000$ candidates are found from photometric data on the SkyMapper telescope, 2618 of which have spectroscopic observations at ANU's 2.3m telescope \citep{dac19}, and for $\sim 500$ stars detailed elemental abundances are measured with higher-resolution spectra taken at larger telescopes such as Magellan, Keck, and VLT \citep{yon21b}. SMSS J200322.54-114203.3 was discovered in the SkyMapper EMP survey and reported in \citet{yon21a}.

Figure \ref{fig:yong} shows the elemental abundance of SMSS J200322.54-114203.3, which has [Fe/H]$= -3.5$, and is 2.3 kpc away with the Galactic halo orbit. This star has a very clear detection of uranium and thorium, and thus it is a so-called actinides boost star. The stars also clearly showed the solar r-process pattern from $Z \sim 60$ to 70. Surprisingly, the observed Zn abundance is very high ([Zn/Fe]$=0.72$), which indicates that the enrichment source was an energetic explosion ($\gtsim 10^{52}$ erg). It also showed normal $\alpha$ enhancement as for the majority of Galactic halo stars, although a normal supernova ($\sim 10^{51}$ erg) gives a range of $\alpha$ enhancement depending on the progenitor mass.
The star is not carbon enhanced ([C/Fe]$<0.07$) but shows nitrogen enhancement ([N/Fe]$=1.07$), which indicates that the progenitor stars were rotating. All of these features at $Z\le 30$ are consistent with the enrichment from a single hypernova.

The observed abundance pattern is compared with two theoretical models. The blue dashed line is for mono-enrichment from a single, magneto-rotational hypernova (MRHN), where the nucleosynthesis yields are obtained combining the iron-core of a magneto-rotational supernovae \citep{nis15} to the envelope of a $25M_\odot$ hypernova. The orange dot-dashed line shows a model of multi-enrichment where a neutron star merger occur in the interstellar matter chemically enriched by previous generations of core-collapse supernovae. The MRHN model gives better fit at and below Zn ($Z\le30$). Both models show overproduction around the 1st peak ($Z\sim40$) and around Ba ($Z=56$), which may be due to the uncertainties in nuclear astrophysics. The MRSN scenario is also consistent with the short timescale of the EMP star formation; due to the required long timescales of neutron star mergers, it is unlikely that this EMP star is enriched by a neutron star merger.

{\bf Future prospects}: This star provided the first observational evidence of the r-process associated with core-collapse supernovae. The r-process site could be in the jets as in the MRHN model, or could also be in the accretion disk around black hole as in the collapsar model \citep{sie19}. According to the chemical evolution models, this event is rare, only one in 1000 supernovae. MRHNe should be observable in future astronomical transient surveys such as with LSST/Rubin, since they eject a significant amount of iron; the ejected iron mass is $0.017M_\odot$ in the $25M_\odot$ model, while $0.33M_\odot$ in the $40M_\odot$ model \citep{yon21a}. It might be possible to detect the signature of Au and Pt production directly in the supernova spectra.

\begin{figure}[t]
\begin{center}
  \includegraphics[width=0.7\textwidth]{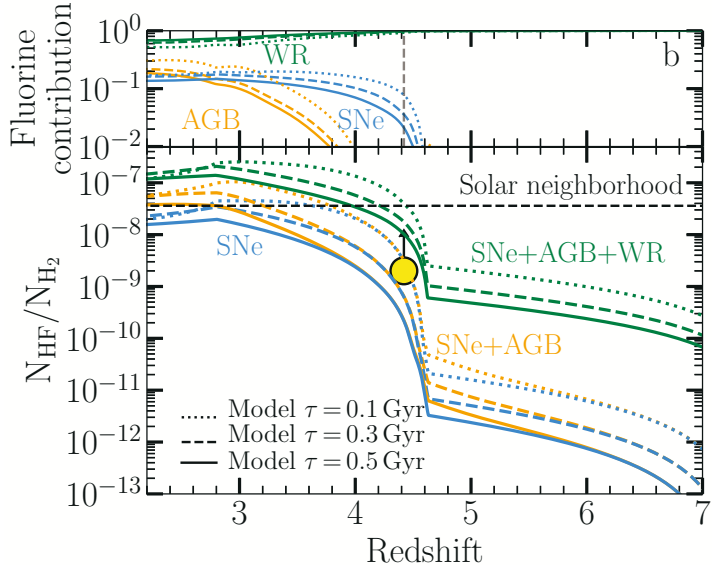}
\caption{The fluorine abundance in the galaxy NGP-190387 at redshift $z=4.4$, comparing with galactic chemical evolution models with only supernovae (blue lines), plus AGB stars (orange lines), plus WR stars (green lines).
}
\label{fig:franco}
\end{center}
\end{figure}

\subsection{Ramp-up of fluorine at high-redshifts}
\label{sec:f}

Fluorine is an intriguing element. The major production site has been debated for 30 years among low-mass AGB stars, stellar winds in massive stars, and the $\nu$-process in core-collapse supernovae \citep{kob11f}. The observational constraints have been obtained for stars, but accessible lines are only in infrared, and thus the sample number was limited. Moreover, there was confusion in the excitation energies and transition probabilities for the HF lines \citep{jon14}. The vibrational-rotational lines get too weak at low metallicities, and the available sample of EMP stars is only for carbon enhanced stars \citep{mur20}, which may not be representative of galactic chemical evolution.
\citet{fra21} opened a new window for constraining fluorine production in the early universe by directly measuring HF abundance in a galaxy at redshift $z=4.4$.

The galaxy NGP-190387 was discovered by the H-ATLAS survey with the Herschel satellite, with the redshift confirmed by the Northern Extended Millimetre Array (NOEMA). It is a gravitationally lensed galaxy with a magnification factor $\mu\simeq5$. The lowest rotational transitions from the HF molecule appeared as an absorption line in the Atacama Large Millimeter/submillimeter Array (ALMA) data. This molecule is very stable and the dominant gas-phase form of fluorine in the ISM.

Figure \ref{fig:franco} shows the fluorine abundance of NGP-190387 (yellow point), comparing with galactic chemical evolution models with various star formation timescales. In the models, it is assumed that star formation took place soon after the reionization, which was boosted probably due to galaxy merger 100 Myr before the observed redshift $z=4.4$. This assumption is based on the properties of other submillimetre galaxies (SMGs) at similar redshifts. The models show rapid chemical enrichment including fluorine, but in order to explain the observed fluorine abundance, the contribution from Wolf-Rayet stars (green lines) is required. The nucleosynthesis yields from \citet{lim18} are adopted in the models, assuming the rotational velocity of 300 km s$^{-1}$ for $13-120M_\odot$ stars.

{\bf Future prospects}: This galaxy provided the first observational evidence for rotating massive stars with initial masses $\gtsim 30M_\odot$ producing a significant amount of fluorine in the early universe.
Since it is not easy to explode these stars with $\nu$-driven mechanism, these stars are likely to remain black holes with masses $\gtsim 20 M_\odot$, which may be detected with gravitational wave missions. 
It would be possible to constrain the formation history further if CNO abundances and/or $^{13}$C isotopic ratio are measured, because the yields depend on the mass of progenitor stars.

\subsection{GCE with rotating massive stars}
\label{sec:iso}

\begin{figure}[t]
\begin{center}
\includegraphics[width=0.95\textwidth]{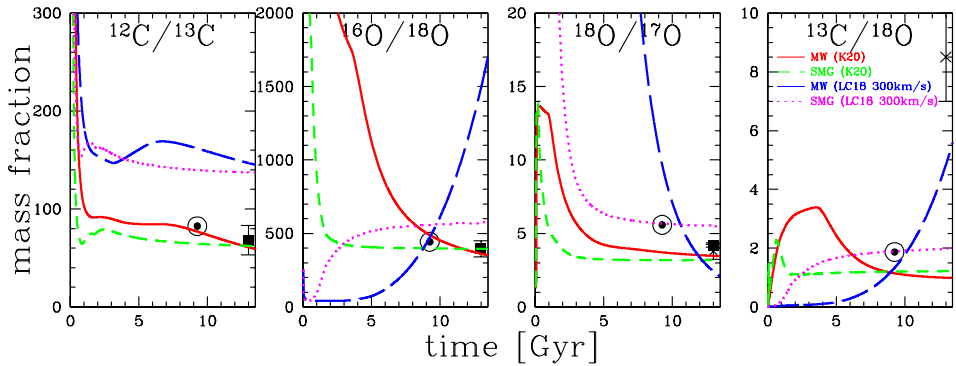}
\vspace*{-1mm}
\caption{\label{fig:iso}
Evolution of isotopic ratios of GCE models for the Milky Way (red solid and blue long-dashed lines) and for a submillimetre galaxy (green short-dashed and magenta dotted lines), which are calculated with the K20 yields (red solid and green short-dashed lines) or the yields from \citet{lim18} with the rotational velocity of 300 km s$^{-1}$ (blue long-dashed and magenta dotted lines).
The solar symbol indicates the solar ratios \citep{asp09}.
See \citet{rom19} for the observational data sources of the ISM (filled squares and crosses).
}
\end{center}
\end{figure}

The evolution of isotopic ratios is shown in Figs.\,17-19 of \citet{kob11agb} and Fig.\,31 of K20 (see also \citealt{rom19}). $^{13}$C and $^{25,26}$Mg are produced from AGB stars ($^{17}$O might be overproduced in our AGB models), while other minor isotopes are more produced from metal-rich massive stars, and thus the ratios between major and minor isotopes (e.g., $^{12}$C/$^{13}$C, $^{16}$O/$^{17,18}$O) generally decrease as a function of time/metallicity. But this evolutionary trend is completely different if rotating massive stars are included.

Figure \ref{fig:iso} shows the evolution of isotopic ratios in the GCE models with and without rotating massive stars, comparing to observational data.
For the Milky Way (red solid and blue long-dashed lines), the adopted star formation history is the same as the solar neighbourhood model in \S \ref{sec:gce}.
For the SMG (green short-dashed and magenta dotted lines), rapid gas accretion and star formation with the timescales of 1 Gyr are assumed, and after $\sim 1.5$ Gyr the star formation is self-regulated with a constant gas fraction of 50\% (see M. Doherty et al., in prep. for more details).
The K20 model (red solid lines) is not bad, well reproducing C and O isotopic ratios in the Sun (the solar symbol) and in the present-day ISM (filled squares), but not the $^{13}$C/$^{18}$O ratio (cross).

The model with rotating massive stars (blue long-dashed lines) gives a better evolutionary track of the $^{13}$C/$^{18}$O ratio, showing a rapid increase in the past 5 Gyrs in the Milky Way. However, $^{12}$C is overproduced at all metallicities, and $^{16}$O is overproduced at high metallicities in this model.
Here the yields from \citet{lim18} with the rotational velocity of 300 km s$^{-1}$ are used. With 150 km s$^{-1}$, the $^{12}$C/$^{13}$C ratio becomes even larger, and the $^{13}$C/$^{18}$O ratio becomes slightly smaller. The model with 150 km s$^{-1}$ is worse than the model with 300 km s$^{-1}$ in order to reproduce the observations, and using an averaged rotational velocity as in \citet{pra18} does not solve this problem.
With these yields set, even excluding stellar rotation, the $^{12}$C/$^{13}$C ratio is systematically higher than in the K20 yields, and the difference may be caused by their choice of nuclear reaction rates, mass-loss, and/or convection.
It is possible that these effects could change the evolution of the $^{18}$O/$^{17}$O ratio.
The overproduction of $^{16}$O is probably due to their choice of remnant masses, and in fact these yields do not reproduce the observed [O/Fe]--[Fe/H] relation.

\begin{figure}[t]
\begin{center}
  \includegraphics[width=0.6\textwidth]{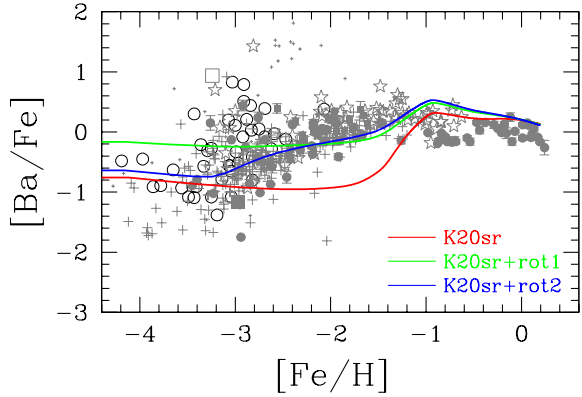}
\vspace*{-2mm}
\caption{The [Ba/Fe]--[Fe/H] relation, comparing with K20 model (red line) and models with rotating massive stars with different rotational velocities (green and blue lines). See K20 for the other data points and model details.}
\label{fig:ba}
\end{center}
\end{figure}

The impact of stellar rotation should be studied from s-process elements, since the weak-s process forms s-process elements from the existing seeds at a much shorter timescale than AGB stars \citep{fri16,lim18}.
For s-process elements, there is some underproduction from [Fe/H] $\sim -3$ to $\sim -1$ in the K20 model, which could be improved in the inhomogeneous enrichment effect in chemodynamical simulations (\S \ref{sec:n}). Alternatively, the models with rotating massive stars can explain the observed average trend (Fig.\,\ref{fig:ba}); in the first model (green line, also plotted in \citealt{yon21b}) the same metallicity-dependent distribution of rotational velocity as in \citet{pra18} is used, while in the second and better model (blue line) 150 and 300 km s$^{-1}$ is assumed for HNe and MRHNe, respectively.
As already noted, the nucleosynthesis yields from \citet{lim18} cannot reproduce the observations of many elements, and thus we use the contributions from stellar envelopes and winds only in addition to the K20 yields (see Kobayashi et al., in prep. for more details).

\vspace*{-2mm}
\section{Chemodynamical evolution of galaxies}
\label{sec:hydro}

Thanks to the development of super computers and numerical techniques, it is now possible to simulate the formation and evolution of galaxies from cosmological initial conditions, i.e., initial perturbation based on $\Lambda$ cold dark matter (CDM) cosmology. Star formation, gas inflow, and outflow in Eq.(\ref{eq:gce}) are not assumed but are, in principle, given by dynamics. Due to the limited resolution, star forming regions in galaxy simulations, supernova ejecta, and active galactic nuclei (AGN) cannot be resolved, and thus it is necessary to model star formation and the subsequent effects -- feedback -- introducing a few parameters. Fortunately, there are many observational constraints, from which it is usually possible to choose a best set of parameters. To ensure this, it is also necessary to run the simulation until the present-day, $z=0$, and reproduce a number of observed relations at various redshifts.

Although hydrodynamics can be calculated with publicly available codes such as Gadget, RAMSES, and AREPO, modelling of baryon physics is the key and is different depending on the simulation teams/runs, such as EAGLE, Illustris, Horizon-AGN, Magneticum, and SIMBA simulations.
These are simulations of a cosmological box with periodic boundary conditions, and contain galaxies with a wide mass range (e.g., $10^{9-12} M_\odot$ in stellar mass) at $z=0$.
In order to study massive galaxies it is necessary to increase the size of the simulation box (e.g., $\gtsim 100$ Mpc), while in order to study internal structures it is necessary to improve the spacial resolution (e.g., $\ltsim 0.5$ kpc).
The box size and resolution are chosen depending on available computational resources.
On the other hand, zoom-in techniques allow us to increase the resolution focusing a particular galaxy,
but this also requires tuning the parameters with the same resolution, comparing to a number of observations in the Milky Way.

For the baryon physics, the first process to calculate is 
{\bf radiative cooling}, and photo-heating by a uniform
and evolving UV background radiation. %\citep{haa96}.
We use a metallicity-dependent
  cooling function computed with the MAPPINGS III
  software %(\citealt{sut93}, 
(see \citealt{kob07} for the details),
and find candidate gas particles that can form stars in a given timestep.
The effect of $\alpha$ enhancement relative to Fe is taken into account using the observed relation in the solar neighborhood.

 Our {\bf star formation} criteria are:
(1)
  converging flow, $(\nabla \cdot \mbox{\boldmath$v$})_i < 0$; (2) rapid
  cooling, $t_{\rm cool} < t_{\rm dyn}$; and (3) Jeans unstable gas, $t_{\rm
    dyn} < t_{\rm sound}$.  The star formation timescale is taken to be
  proportional to the dynamical timescale ($t_{\rm sf} \equiv
  \frac{1}{c_*}t_{\rm dyn}$), where $c_*$ is a star formation timescale
  parameter. 
The parameter $c_*$ basically determines when to form stars following cosmological accretion, and has a great impact on the size--mass relation of galaxies (Fig.\,4 of \citealt{kob05}).
We found that it is better not to include a fixed density criterion.
Note that based on smaller-scale simulations,
more sophisticated analytic formula are proposed
including the effects of turbulence and magnetic fields.

A fractional part of the mass of the gas
  particle turns into a new star particle \citep[see][for the details]{kob07}.
Note that an individual
  star particle has a typical mass of $\sim 10^{5-7} M_\odot$, i.e.~it does not
  represent a single star but an association of many stars.  The masses of
  the stars associated with each star particle are distributed according to an
  IMF.  We adopt Kroupa IMF, which is assumed to be
  independent of time and metallicity, and 
 limit the IMF to the mass range
  $0.01M_\odot \le m \le 120M_\odot$.
This assumption is probably valid for the resolution down to $\sim 10^4 M_\odot$.

We then follow the evolution of the star particle, as a simple stellar population, which is defined as a single generation of coeval and chemically homogeneous stars of various masses, i.e. it consists of a number of stars with various masses but the same age and chemical composition \citep{kob04}.
The production of each element $i$ from the star particle (with an initial mass of $m_*^0$) is calculated using a very similar equation as Eq.(\ref{eq:gce}):
\begin{equation}\label{eq:e-z}
E_Z(t)= m_*^0 \left[ E_{\rm SW}+E_{\rm SNcc}+E_{\rm SNIa}+E_{\rm NSM} \right] .
\end{equation}
Similarly, the energy production from the star particle is:
\begin{equation}\label{eq:e-e}
E_e(t) = m_*^0 \left[ e_{e,{\rm SW}}{\cal{R}}_{\rm SW}(t)+e_{e,{\rm SNcc}}{\cal{R}}_{\rm SNcc}(t)+e_{e,{\rm SNIa}}{\cal{R}}_{\rm SNIa}(t) \right]
\end{equation}
which includes the event rates of SWs (${\cal{R}}_{\rm SW}$), core-collapse supernovae (${\cal{R}}_{\rm SNcc}$), and SNe Ia (${\cal{R}}_{\rm SNIa}$), and the energy per event (in erg) for SWs: $e_{e,{\rm SW}}=0.2 \times 10^{51} ({Z}/{Z_\odot})^{0.8}$ for $>8M_\odot$ stars, Type II supernovae: $e_{e,{\rm SNII}}=1 \times 10^{51}$, hypernovae: $e_{e,{\rm HN}}=10,10,20,30 \times 10^{51}$ for $20,25,30,40 M_\odot$ stars, respectively, and for SNe Ia: $e_{e,{\rm SNII}}=1.3 \times 10^{51}$.

\begin{figure}[t]
\begin{center}
\includegraphics[width=8cm]{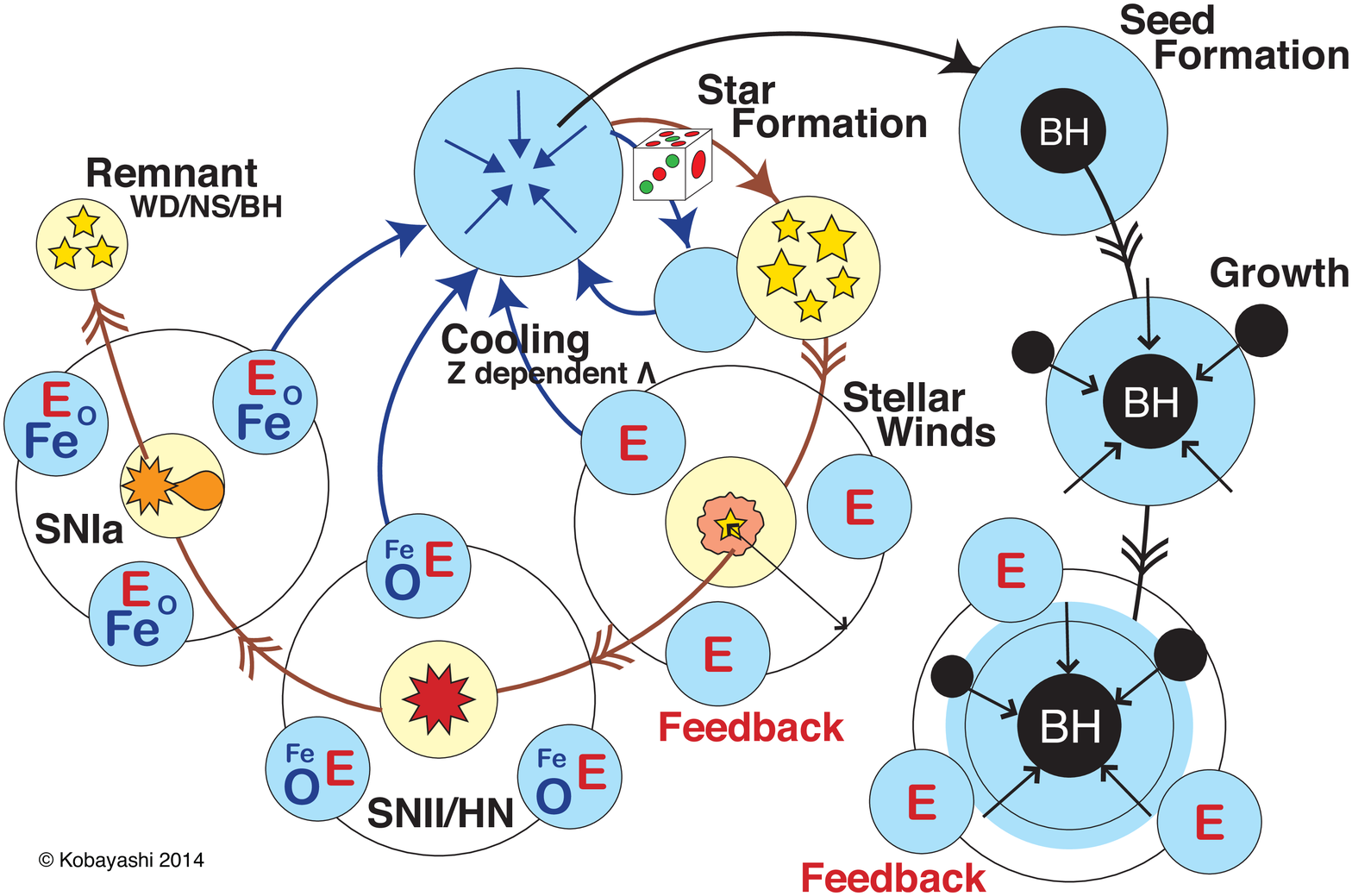}
\caption{\label{fig:hydro}
Schematic view of chemodynamical evolution of galaxies.
}
\end{center}
\end{figure}

We distribute this {\bf feedback} metal mass and energy from a star particle $j$
to a fixed number of neighbour gas particles $\ell$, $N_{\rm FB}$, weighted by the smoothing kernel as:
\begin{equation}\label{eq:fb-z}
\frac{d(Z_{i,k}\,m_{{\rm g},k})}{dt}= \sum_j \left[W_{kj} E_{Z,j}(t) / \sum_\ell^{N_{\rm FB}} W_{\ell j} \right]
\end{equation}
and
\begin{equation}\label{eq:fb-e}
\frac{du_k}{dt}= (1-f_{\rm kin}) \sum_j \left[W_{kj} E_{e,j}(t) / \sum_\ell^{N_{\rm FB}} W_{\ell j} \right]
\end{equation}
where $u_k$ denotes the internal energy of a gas particle $k$, and $f_{\rm kin}$ denotes the kinetic energy fraction, i.e., the fraction of energy that is distributed as a random velocity kick to the gas particle.
Note that in the first sum the number of neighbour star particles is not fixed to $N_{\rm FB}$.
To calculate these equations, the feedback neighbour search needs to be done twice in order to ensure proper
mass and energy conservation; first to compute the sum of weights for the
normalization, and a second time for the actual distribution.

The parameter $N_{\rm FB}$ determines the average energy distributed to gas particles; a large value of $N_{\rm FB}$ leads to inefficient feedback as the ejected energy radiatively cools away shortly, while a small value of $N_{\rm FB}$ results in a only small fraction of matter affected by feedback.
Alternatively, with good resolution in zoom-in simulations, feedback neighbors could be chosen within a fixed radius ($r_{\rm FB}$), which affects a number of observations including the MDF (Fig.\,12 of \citealt{kob11mw}).
The parameter $f_{\rm kin}$ has a great impact on radial metallicity gradients in galaxies, and $f_{\rm kin}=0$, i.e., purely thermal feedback, gives the best match to the observations (Fig.\,14 of \citealt{kob04}).
Note that various feedback methods are proposed such as
the stochastic feedback \citep[][used for EAGLE]{dal08} and the mechanical feedback \citep[][used for FIRE]{hop18}.

Eq.(\ref{eq:fb-z}) gives positive feedback that enhances radiative cooling, while Eq.(\ref{eq:fb-e}) gives negative feedback that suppresses star formation.
The mass ejection in Eq.(\ref{eq:fb-z}) never becomes zero, while the energy production in Eq.(\ref{eq:fb-e}) becomes small after 35 Myrs.
Therefore, it is not easy to control star formation histories depending on the mass/size of galaxies within this framework.
In particular, it is not possible to quench star formation in massive galaxies, since low-mass stars keep returning their envelope mass into the ISM for a long timescale, which will cool and keep forming stars.

Therefore, in order to reproduce observed properties of massive galaxies, additional feedback was required, and the discovery of the co-evolution of super-massive black holes and host galaxies provided a solution -- {\bf AGN feedback}.
Modelling of AGN feedback consists of (1) seed formation, (2) growth by mergers and gas accretion, and (3) thermal and/or kinetic feedback (see \citealt{tay14} for the details).
In summary, we introduced a seeding model where seed black holes originate from the formation of the first stars, which is different from the `standard' model by \citet{spr05agn} and from most large-scale hydrodynamical simulations. %\cite{dub12,vog14,sch14,pil18,dav19}.
In our AGN model, 
seed black holes are generated if the metallicity of the gas cloud is zero ($Z=0$) and the density is higher than a threshold density ($\rho_{\rm g} > \rho_{\rm c}$). 
The growth of black holes is the same as in other cosmological simulations, and is calculated with Bondi-Hoyle accretion, swallowing of ambient gas particles, and merging with other black holes.
Since we start from relatively small seeds, the black hole growth is driven by mergers at $z\gtsim 3$ (Fig.\,2 of \citealt{tay14}).
On a very small scale, it is not easy to merge two black holes, and an additional time-delay is applied in more recent simulations (P. Taylor et al. in prep.).

Proportional to the accretion rate, thermal energy is distributed to the surrounding gas, which is also the same as in many other simulations.
In more recent simulations, non-isotropic distribution of feedback area is used to mimic the small-scale jet \citep[e.g.,][]{dav19}.
There are a few parameters in our chemodynamical simulation code, but \citet{tay14} constrained the model parameters from observations, and determined the best parameter set: $\alpha=1$, $\epsilon_{\,\rm f}=0.25$, $M_{\rm seed}=h^{-1} 10^3M_\odot$, and $\rho_{\rm c}=0.1 h^2 m_{\rm H}\,\textrm{cm}^{-3}$.
Our black holes seeds are indeed the debris of the first stars although we explored a parameter space of $10^{1-5}M_\odot$.
This is not the only one channel for seeding, and the direct collapse of primordial gas and/or the collapse of dense stellar clusters ($\sim 10^5M_\odot$, \citealt{woo19}) are rarer but should be included in larger volume simulations.
Nonetheless, our model can successfully drive large-scale galactic winds from massive galaxies \citep{tay15b} and can reproduce many observations of galaxies with stellar masses of $\sim 10^{9-12}M_\odot$ \citep{tay15a,tay16,tay17}.
The movie of our fiducial run is available at \url{https://www.youtube.com/watch?v=jk5bLrVI8Tw}

\subsection{Extra-galactic archaeology}
\label{sec:n}

\begin{figure}[t]
\begin{center}
  \includegraphics[width=0.35\textwidth]{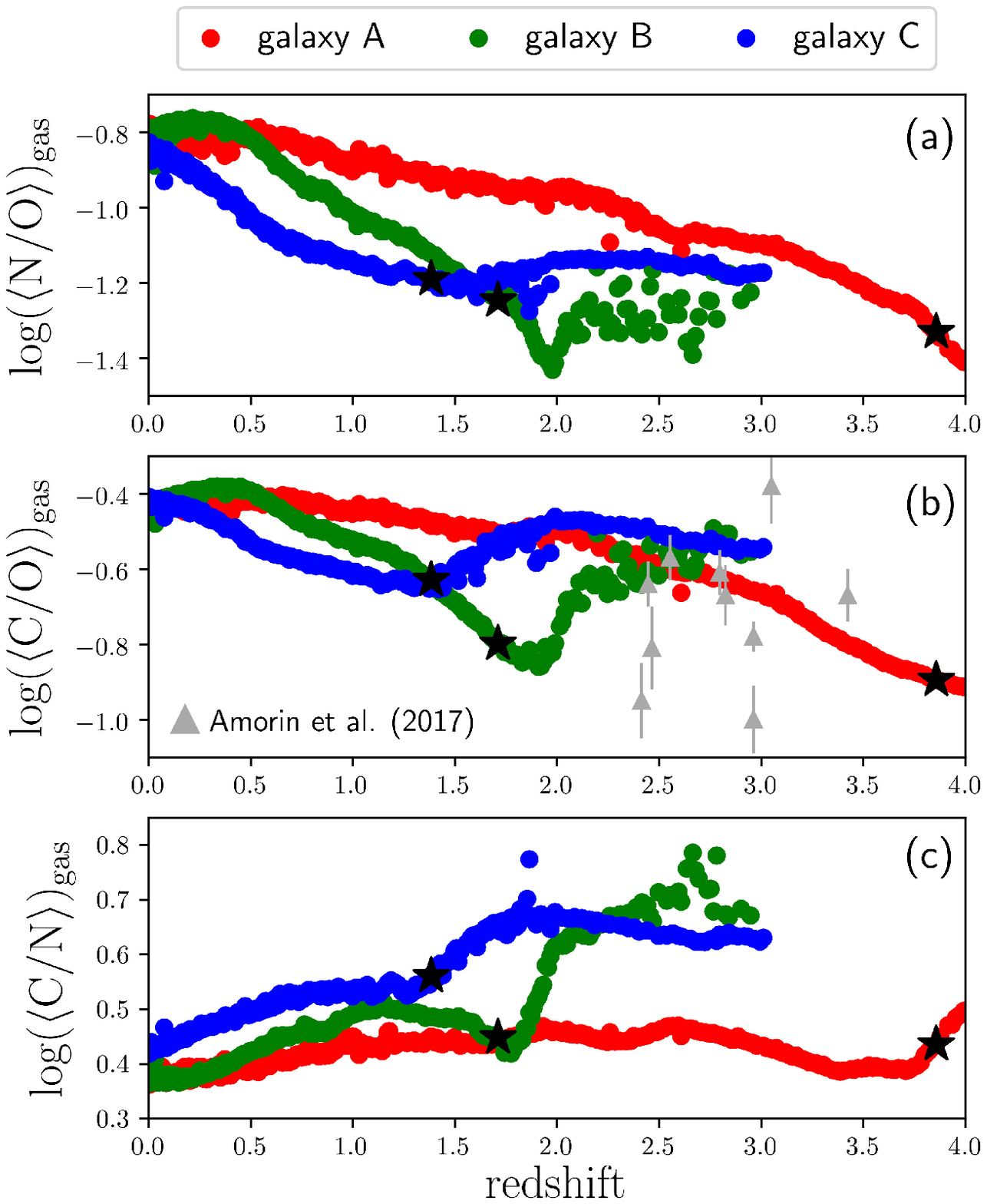}
  \includegraphics[width=0.60\textwidth]{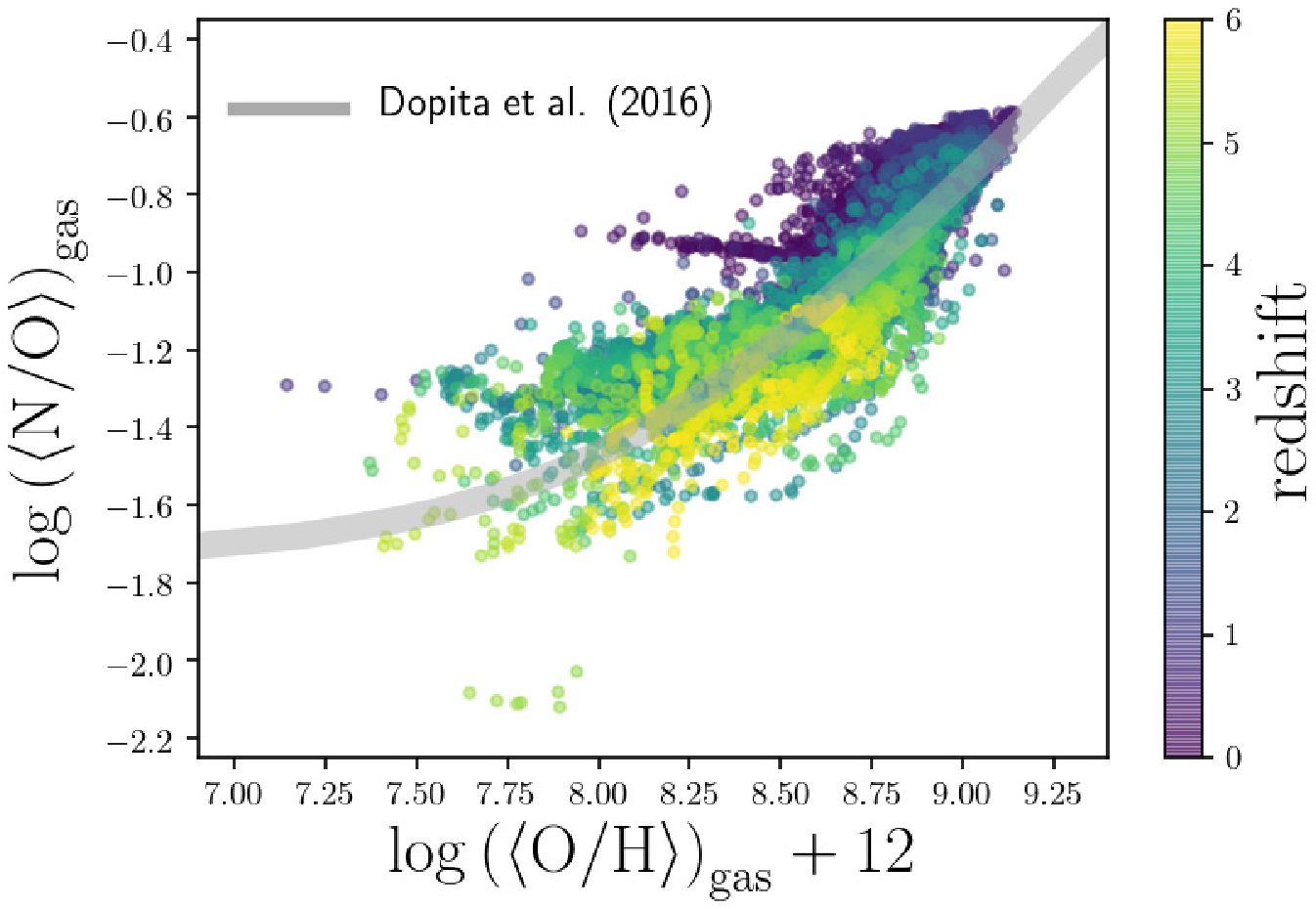}
\caption{(Left) Evolution of CNO abundance ratios in simulated disk galaxies with different formation timescales, taken in a cosmological simulation.
(Right) Evolution of the N/O--O/H relation of simulated galaxies in a cosmological simulation. The gray bar indicates the compilation of observational data.
}
\label{fig:cno}
\end{center}
\end{figure}

Elemental abundances and isotopic ratios provide additional constrains on the timescales of formation and evolution of galaxies. This extra-galactic archaeology will be possible in the near future with the James Webb Space Telescope (JWST), the wavelength coverage of which will allow us to measure CNO abundance simultaneously \citep{vin18a}.
 At high redshifts, metallicities have been measured mainly with emission lines in star-forming galaxies, where Fe abundance is not accessible. Instead of $\alpha$/Fe ratios in Galactic archaeology, CNO abundances can be used for extra-galactic archaeology. 

The left panel of Figure \ref{fig:cno} shows theoretical predictions of CNO abundance ratios for simulated disk galaxies that have different formation timescales. These galaxies are chosen from a cosmological simulation, which is run by a Gadget-3 based code that includes detailed chemical enrichment \citep{kob07}. In the simulation, C is mainly produced from low-mass stars ($\ltsim 4M_\odot$), N is from intermediate-mass stars ($\gtsim 4M_\odot$) as a primary process \citep{kob11agb}, and O is from massive stars ($\gtsim 13M_\odot$). C and N are also produced by massive stars; the N yield depends on the metallicity as a secondary process, and can be greatly enhanced by stellar rotation (as for F in \S \ref{sec:f}).

In the nearby universe, the N/O--O/H relation is known for stellar and ISM abundances, which shows a plateau ($\sim -1.6$) at low metallicities and a rapid increase toward higher metallicities. This was interpreted as the necessity of rotating massive stars by \citet{chi06}. However, this should be studied with hydrodynamical simulations including detailed chemical enrichment, and \citet{kob14mw} first showed the N/O--O/H relation in a chemodynamical simulation. \citet{vin18b} showed that both the global relation, which is obtained for average abundances of the entire galaxies, and the local relation, which is obtained for spatially resolved abundances from IFU data, can be reproduced by the inhomogeneous enrichment from AGB stars. Since N yield increases at higher metallicities, the global relation originates from the mass--metallicity relation of galaxies, while the local relation is caused by radial metallicity gradients within galaxies. 
Moreover, the right panel of Figure \ref{fig:cno} shows a theoretical prediction on the time evolution of the N/O--O/H relation, where galaxies evolve along the relation. Recent observation with KMOS (KLEVER survey) confirmed a near redshift-invariant N/O-O/H relation \citep{hay22}. 

{\bf Future prospects}:
It is possible to reproduce the observed N/O--O/H relation only with AGB stars and supernovae in our chemodynamical simulations. If we include N production from WR stars, the N/O plateau value at low metallicities becomes too high compared with observations.
The enrichment from the first stars is not included either, which can be significantly different because very massive stars ($\ge 140M_\odot$) explode as pair-instability supernovae without leaving any remnants, or massive stars ($\sim 25M_\odot$) explode as failed supernovae that leave relatively more massive black holes ($\sim 5M_\odot$). The former is based on theoretical prediction (but no observational evidence yet) and the latter is inferred from observational results \citep{nom13}.
It would be very useful if other elemental abundances (e.g., Ne, S) are measured in order to determine the chemical enrichment sources and constrain the formation histories of galaxies.

\subsection{Galactic archaeology}
\label{sec:mw}

Inhomogeneous enrichment in chemodynamical simulations leads to a paradigm shift on the chemical evolution of galaxies.
As in a real galaxy, i) the star formation history is not a simple function of radius, ii) the ISM is not homogeneous at any time, and iii) stars migrate \citep{vin20}, which are fundamentally different from one-zone or multi-zones GCE models.
As a consequence, (1) there is no age--metallicity relation, namely for stars formed in merging galaxies. It is possible to form extremely metal-poor stars from accretion of nearly primordial gas, or in isolated chemically-primitive regions, at a later time.
(2) Enrichment sources with long time-delays such as AGB stars, SNe Ia, and NSMs can appear at low metallicities. This is the reason why the N/O plateau is caused by AGB stars in \S \ref{sec:n}. However, this effect is not enough to reproduce the observed [Eu/(O,Fe)] ratios only with NSMs \citep{hay19}.
(3) There is a significant scatter in elemental abundance ratios at a given time/metallicity, as shown in Figs.\,16-18 of \citet{kob11mw}.

Figure \ref{fig:mwxfe} shows the frequency distribution of elemental abundance ratios in our chemodynamical zoom-in simulation of a Milky Way-type galaxy ($r\le100$ kpc). The colour indicates the number of stars in logarithmic scale. In order to compare the scatter, the range of y-axis is set to the same, from $-1.5$ to $+0.5$ for the elements except for He and Li.
In the simulation we know the ages and kinematics of star particles as well as the locations within the galaxy, which are used in the following summary:

\begin{figure*}[t]
\begin{center}
  \includegraphics[width=0.95\textwidth]{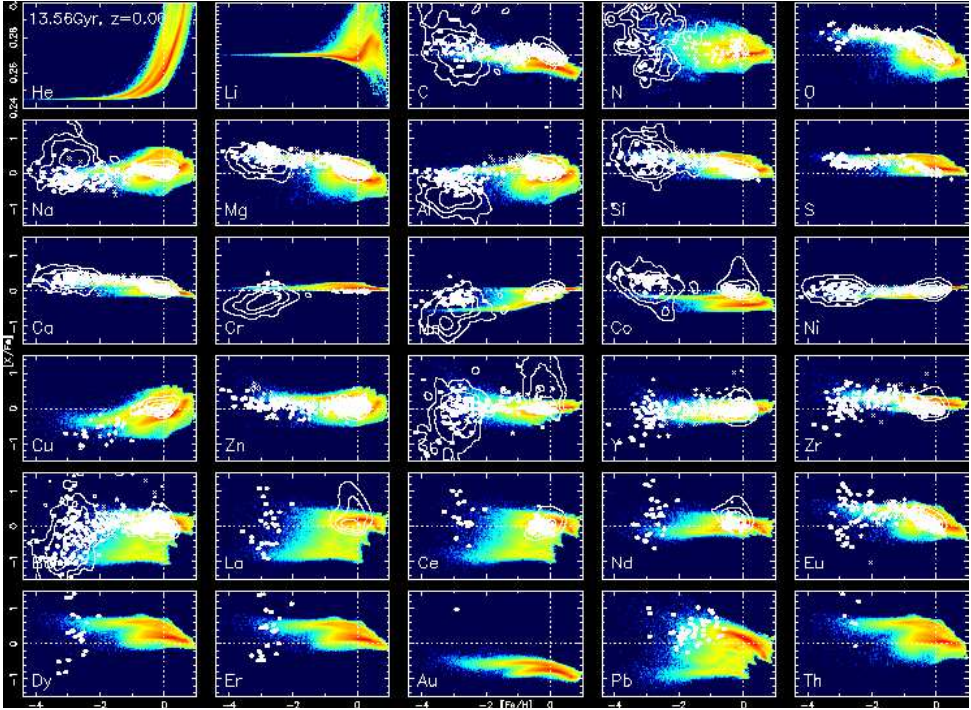}
\vspace*{-1mm}
\caption{Frequency distribution of elemental abundance ratios in a chemodynamical zoom-in simulation of a Milky Way-type galaxy.
The filled circles are observational data from high-resolution spectra, and the metal-rich and metal-poor contours show the HERMES-GALAH survey DR3 \citep{bud21} and the SkyMapper EMP survey \citep{yon21b}, respectively.
}
\label{fig:mwxfe}
\end{center}
\end{figure*}

\begin{itemize}
\item The Big Bang nucleosynthesis gives $Y=0.2449$.
He is also synthesized during stellar evolution and the He abundance increases up to $Y \sim 0.33$ showing a tight function of metallicity (see also \citealt{vin19}).
\item The initial Li abundance is set to be a theoretical value of the Big Bang nucleosynthesis, $A({\rm Li})=2.75$, although it is higher than the observed value of EMP stars.
The Li abundance increases by the production from AGB stars, but decreases at super-solar metallicities due to the destruction in AGB stars.
There is a large scatter from $A({\rm Li})\sim2.6$ to $0.3$ at [Fe/H] $\gtsim -1$.
\item The average [C/Fe] ratio is around $\sim 0$. There is only a small number of C-enhanced EMP stars with [C/Fe] $\ge0.7$ due to the inhomogeneous enrichment from AGB stars in our model. This number is much smaller than observed because faint supernovae are not included.
\item The [N/Fe] ratio shows a large scatter of $\pm0.5$ dex even at [Fe/H] $\sim -3$ due to the inhomogeneous enrichment, and it is possible to reproduce the observational data only with AGB stars, without rotating massive stars.
\item The $\alpha$ elements, O, Mg, Si, S, and Ca, show similar trends, a plateau with a small scatter at [Fe/H] $\ltsim -1$, and then bimodal decreasing trends both caused by the delayed Fe enrichment from SNe Ia. The high-$\alpha$ stars tend to have high $v/\sigma$ and old ages, forming the thick disk. On the other hand, the low $\alpha$ stars are rotationally supported and relatively younger, belonging to the thin disk.
The thick disk reaches only [Fe/H] $\sim 0$, while thin disk expends to [Fe/H] $\sim 0.2$. The stars with [Fe/H] $>0.2$ in this figure are located in the bulge region ($r\le 1$ kpc in this paper).
\item The odd-$Z$ elements, Na, Al, and Cu, show bimodal increasing trends from [Fe/H] $\sim -2$ to higher metallicities, which are predominantly caused by the metallicity dependent yields of core-collapse supernovae.
The higher [(Na,Al,Cu)/Fe] trend is composed of the high $\alpha$ stars in the thick disk, while the lower [(Na,Al,Cu)/Fe] trend is made of the low $\alpha$ stars in the thin disk; these differences are caused by SNe Ia.
Moreover, [Na/Fe] shows a greater increase with a steeper slope for the thick disk than that for the thin disk, which is due to an additional contribution from AGB stars. Such a difference in the slopes is not seen for Al and Cu.
\item The [Cr/Fe] ratio is almost constant at $\sim 0$ over all ranges of metallicities. In observations, EMP stars tend to show [Cr/Fe] $<0$ values affected by the NLTE effect, but Cr II observations show [Cr/Fe] $\sim 0$ (K06). No decrease of [Cr/Fe] is caused by SNe Ia either if we use the latest SN Ia yields \citep{kob20ia}.
\item In contrast to the [$\alpha$/Fe] ratios, [Mn/Fe] shows a plateau value of $\sim -0.5$ at [Fe/H] $\ltsim -1$, and then bimodal increasing trends because Ch-mass SNe Ia produce more Mn than Fe (K06). Unlike one-zone models, [Mn/Fe] does not reach a very high value ($\sim +0.5$) in chemodynamical simulations because not all SN Ia progenitors have high metallicities due to the inhomogeneity. The inclusion of sub-Ch mass SNe Ia would worsen the situation.
\item Co and Zn can be enhanced by higher energy explosions (K06). There are two plateaus with [Zn/Fe] $\sim 0.2$ for the thick disk and [Zn/Fe] $\sim 0$ for the thin disk. At [Fe/H] $>0.2$, [Zn/Fe] shows a rapid increase, which is seen only for the bulge stars in our model.
Very similar features are seen also for Co, although Co is under-produced overall, which is due to the lack of multi-dimensional effect in the nucleosynthesis yields \citep{sne16}.
\item The first-peak s-process elements (Sr, Y, and Zr) are produced by ECSNe and AGB stars. Sr and Y do not show a bimodality, while Zr shows a very similar trend as for the $\alpha$ elements.
\item The second-peak s-process elements (Ba, La, and Ce) show a very large scatter of $1.5$ dex especially at low metallicities. These elements are first produced by MRSNe/MRHNe in our model with a floor value of [(Ba,La,Ce)/Fe] $\sim -1$, and are greatly enhanced by AGB stars. The contribution from AGB stars appears from [Fe/H] $\sim -2$;  [(Ba,La,Ce)/Fe] ratios reach $\sim 0$ at [Fe/H] $\sim -1$, peaks at [Fe/H] $\sim 0$, then decrease due to SNe Ia. The decrease is significant in the bulge stars at [Fe/H] $>0.2$.
\item The r-process elements (Nd, Eu, Dy, Er, Au, and Th in the figure) do show a bimodality in the model with MRSNe/MRHNe. The average [Nd/Fe] is about $\sim 0$, while Eu, Dy, Er, and Th behave very similar to the $\alpha$ elements.
As discussed in \S \ref{sec:au}, Au is underabundant overall, but the thick disk stars are systematically more gold-rich than for the thin disk stars!
\item Pb is one of the third-peak s-process elements, and the distribution at low metallicities are similar to the second-peak s-process elements. However, [Pb/Fe] shows a steeper decrease from [Fe/H] $\sim -1$ to higher metallicities, due to the metallicity dependence; Pb is produced more from metal-poor AGB stars.
\end{itemize}

Compared with observational data, the scatter of s-process elements seem too large around [Fe/H] $\sim -1$, which also supports the weak-s process rotating massive stars.
The elemental abundance distributions depend on the locations within galaxies, as shown in Fig.\,1 of \citet{kob16mw} for a different chemodynamical simulation.
These distributions should be compared with non-biased, homogeneous dataset of observational data and the comparison to APOGEE DR16 was shown in \citet{vin20}.
More analysis will be shown in Kobayashi (2022, in prep).
The movie of our fiducial run is available at \url{https://star.herts.ac.uk/~chiaki/works/Aq-C-5-kro2.mpg}

\vspace*{-2mm}
\section{Conclusions and Discussion}

Thanks to the long-term collaborations between nuclear and astrophysics, we have good understanding on the origin of elements (except for the elements around Ti and a few neutron-capture elements such as Au).
Inhomogeneous enrichment is extremely important for interpreting the elemental abundance trends (\S \ref{sec:mw}). It can reproduce the observed N/O--O/H relation only with AGB stars and supernovae (\S \ref{sec:n}), but not the observed r-process abundances only with NSMs; an r-process associated with core-collapse supernovae such as magneto-rotational hypernovae is required, although the explosion mechanism is unknown (\S \ref{sec:au}).
It is necessary to run chemodynamical simulations from cosmological initial conditions, including detailed chemical enrichment.
Theoretical predictions depend on input stellar physics, and the effects of stellar mass loss due to rotation and/or binary interaction should be investigated further.
The importance of WR stars is indicated by the high F abundance in $z\sim 4.4$ galaxy (\S \ref{sec:f}), as well as for some isotopic ratios and Ba and Pb abundances around [Fe/H] $\sim -1$ in the Milky Way (\S \ref{sec:iso}).
If WR stars are producing heavy elements on a very short timescale, it might be hard to find very metal-poor or Population III (and dust-free) galaxies at very high redshifts even with JWST.
Finally, mass loss from these stars will change the dust production as well, and thus it is also important to calculate element-by-element dust formation, growth, and destruction, as well as the detailed chemical enrichment.

Galactic archaeology is a powerful approach for reconstructing the formation history of the Milky Way and its satellite galaxies.
APOGEE and HERMES-GALAH surveys have provided homogeneous datasets of many elemental abundances that can be statistically compared with chemodynamical simulations. Future surveys with WEAVE and 4MOST will provide more.
Having said that, the number of EMP stars will not be increased so much in these surveys, and a target survey such as the SkyMapper EMP survey is also needed, in particular for constraining the early chemical enrichment from the first stars.

Reflecting the difference in the formation timescale, elemental abundances depend on the location within galaxies.
Although this dependence has been explored toward the Galactic bulge by APOGEE, the dependence at the outer disk is still unknown, which requires 8m-class telescopes such as the PFS on Subaru telescope. Despite the limited spectral resolution of the PFS, $\alpha$/Fe and a small number of elements will be available.
The PFS will also be able to explore the $\alpha$/Fe bimodality in M31; it is not yet known if M31 has a similar $\alpha$/Fe dichotomy or not.

The next step will be to apply the Galactic archaeology approach to external or distant galaxies. Although it became possible to map metallicity, some elemental abundances, and kinematics within galaxies with IFU, the sample and spacial resolution are still limited even with JWST.
Integrated physical quantities over galaxies, or stacked quantities at a given mass bin, will also be useful, which can be done with the same instruments developed for Galactic archaeology (although optimal spectral resolutions and wavelength coverages are different).
In addition, ALMA has opened a new window for elemental abundances and isotopic ratios in high-redshift galaxies.

It is very important to understand stellar physics when these observational data are translated into the formation timescales or physical processes of galaxies.
For example, the famous [$\alpha$/Fe] ratio is not a perfect clock.
Analysing low-$\alpha$ EMP stars, \citet{kob14} summarized various reasons that cause low $\alpha$/Fe ratios: (1) SNe Ia, (2) less-massive core-collapse supernovae ($\ltsim 20M_\odot$), which become more important with a low star formation rate (3) hypernovae, although the majority of hypernovae are expected to give normal [$\alpha$/Fe] ratios ($\sim 0.4$), and (4) pair-instability supernovae, which could be very important in the early universe.
Therefore, low-$\alpha$ stars are not necessarily enriched by SNe Ia in a system with a long formation timescale. 
It is necessary to also use other elemental abundances (namely, Mn and neutron-capture elements) or isotopic ratios, with higher resolution ($>40,000$) multi-object spectroscopy on 8m class telescopes (e.g., cancelled WFMOS or planned MSE).

\vspace*{-1mm}

\acknowledgement
I thank my collaborators, D. Yong, M. Franco, F. Vincenzo, P. Taylor, A. Karakas, M. Lugaro, N. Tominaga, S.-C. Leung, M. Ishigaki, and K. Nomoto, and V. Springel for providing Gadget-3.
I acknowledge funding from the UK Science and Technology Facility Council through grant ST/M000958/1, ST/R000905/1, ST/V000632/1. 
The work was also funded by a Leverhulme Trust Research Project Grant on ``Birth of Elements''.

\end{document}